\documentstyle[12pt]{article}
\input amssym.def \input amssym

\newcommand{\R}{{\Bbb R}}
\newcommand{\be}[1]{\begin{equation}\label{#1}}
\newcommand{\ee}{\end{equation}}
\newcommand{\beq}[1]{\begin{eqnarray}\label{#1}}
\newcommand{\eeq}{\end{eqnarray}}
\newcommand{\beqo}{\begin{eqnarray*}}
\newcommand{\eeqo}{\end{eqnarray*}}
\newcommand{\li}[1]{\left#1}
\newcommand{\re}[1]{\right#1}
\newcommand{\ez}{\setcounter{equation}{0}\hfill$\mbox{}$\vspace{-1.5em}}
\newcommand{\half}{\frac{1}{2}}

\newcommand{\suml}{\sum\limits}
\newcommand{\qmbox}[1]{\quad\mbox{#1}\quad}
\newcommand{\const}{{\rm const}}
\newcommand{\D}{\partial}
\newcommand{\ga}{\alpha}
\newcommand{\gb}{\beta}
\newcommand{\ba}{\begin{array}}
\newcommand{\ea}{\end{array}}
\newcommand{\dddot}[1]{\ddot{#1}\dot{\hspace{-1.4ex}#1}}
\newcommand{\rfg}[1]{{\rm(\ref{#1})}}

\begin{document}

\title{\bf The de Sitter space-time as attractor
 solution in higher order gravity}
\author{Sabine Kluske\\Universit\"at Potsdam,\\
 Mathematisch-naturwissensch.
Fakult\"at,\\D-14415 Potsdam, Germany}
\date{}

\maketitle

\section{Introduction} %1.
\ez

Hermann Weyl \cite{1} was the first {}to think about
gravitational field
equations of higher than second order. 1951 Buchdahl \cite{2}
gave a general
method for deriving the field equation from a Lagrangian,
written as an
arbitrary invariant of the curvature tensor and its covariant
derivatives up
{}to an arbitrarily high order. He studied the example
$R\square R$. $\square$
denotes the D'Alembertian. This example leads {}to a field
equation of sixth
order. Other generalisations of the Lagrangian are $R^k$,
$R^k\square R$,
\cite{2.5}, $R\square^k R$ and $R\frac{1}{1-\square}R$. The
example $R+aR^2+b
R\square R$ correspondents for finely tuned initial conditions
{}to a cosmological
solution with double inflation. The attempt {}to find more
typical cosmological
solutions with double inflations led {}to $R^k\square R$
\cite{3}, \cite{4}.
This Lagrangian gives unfortunately a theory with unstable
weak field
behaviour. The effort {}to connect the gravitation theory with
quantum theory
leads {}to studies about the Lagrangian
$R\frac{1}{1-\square}R$, \cite{5}. This
Lagrangian can be approximat at the $k$-th step by the sum
\be									{11}
\sum_{i=0}^k R \square^i R~.
\ee
This gives a $2k+4$-th order field equation. The Lagrangian
$c_0R+c_1R\square
R+c_2R\square\square R$ is discussed in \cite{6}. In this
paper we will deduce
the stability properties of the de Sitter solution for $R^k$,
$R\square^kR$
and $\suml_{i=0}^kc_iR\square^iR$.

\section{The field equation} %2.
\ez

The Lagrangian $L=F(R,\square R,\ldots,{\square}^k R)$ leads
{}to a field
equation of $4k+4$-th
\pagebreak
order in general:
\beq									  {21}
0 & = & GR^{ij} - \half Fg^{ij} - G^{;ij} + g^{ij}\square G +
\nonumber\\
  &   & \hspace{2cm} + \sum_{A=1}^k\half g^{ij}
		       \li[F_A(\square^{A-1}R)^{;k}\re]_{;k}
                     - F_A^{;(i}(\square^{A-1}R)^{;j)}~.
\eeq
The abbreviations are:
\be									  {22}
F_A  :=  \sum_{j=A}^k \square^{j-A}\frac{\D F}{\D\square^jR}
\ee
and
\be									  {23}
G  :=  F_0~.
\ee
The operator $\square$ denotes the D'Alembertian, ``," the
partial derivation and
``;" the covariant derivation. For the $D$-dimensional $(D=n+1
\ge 2)$ de
Sitter space-time with ($H\neq 0$) we use
\be									  {24}
ds^2  =  dt^2 - e^{2Ht}\sum_{i=1}^n (dx^i)^2.
\ee
The relation between $H$ and $R$ is
\be									{24.5}
R=-n(n+1)H^2.
\ee
We restrict $R$ {}to the interval $R<0$ subsequently. Other
important relations
are
\be									  {25}
R^{ij}  =  \frac{R}{n+1}g^{ij}
\ee
and
\be									  {26}
\square^k R  =  0 \qmbox{for} k>0~.
\ee
We get the field equation:
\beq									  {27}
0 & = & GR^{ij} - \half Fg^{ij} \nonumber\\
  & = & g^{ij}\li(\frac{1}{n+1}RG - \half F\re)
\eeq
for the de Sitter space-time.

If we choose the Lagrangian $(-R)^u$ with $u\in\R$ the
$D$-dimensional de
Sitter space-time satisfies the field equation iff
$u=\frac{n+1}{2}=\frac{D}{2}$.

An other important Lagrangian is $R\square^k R$ for $k>0$. For
this we get
\be									  {28}
F  =  R \square^k R~,\qquad G  =  2 \square^k R
\ee
and the solubility condition
\be									  {29}
R \square^k R  =  \frac{u}{n+1} R \square^k R
\ee
for the field equation. It is automatically satisfied because
of equation
\rfg{26}

The $D$-dimensional ($D>2$) de Sitter space-time is an exact
solution of the
field equation following from the Lagrangian $R\square^k R$
iff $D\neq4$ and
$k>0$.

\section{The attractor property of the de Sitter space-time}
%3.
\ez

We will examine the attractor property of the de Sitter
space-time in the set
of the spatially flat Friedmann-Robertson-Walker model (FRW
model). We use the
metric
\be									  {31}
ds^2  =  dt^2 - e^{2\ga(t)} \sum_{i=1}^n (dx^i)^2
\ee
for the spatial flat FRW model. For $\ga(t)=Ht$, $H>0$ we get
the de Sitter
space-time metric. {}to find the dynamically behaviour in the
neighbourhood of
the de Sitter space-time we make the ansatz
\be									  {32}
\dot\ga(t)  =  H + \gb(t)
\ee
for the linearisation of the field equation. This is
justified, because the
field equation does not depend on $\ga$ itself, but on its
derivatives only.

We say, the de Sitter space-time is an attractor solution of
the differential
equation if the solutions $\ga(t)$ of the around the de Sitter
space-time
linearized differential equation satisfies
\be									  {33}
\lim_{t\to\infty}\frac{\ga(t)}{t}  =  \widetilde H  =
\const~.
\ee
It is enough {}to discuss the special de Sitter space-time
with $H=1$, because
homothetic and coordinate transformations transfer de Sitter
space-time of the
same dimension into each other.

\section{The Lagrangian F=($-$R)${\bf^u}$} %4.
\ez

We will examine the attractor property of the de Sitter
space-time in the set
of the FRW models for the Lagrangian $(-R)^u$ with
$2u=D=n+1>2$. From this
Lagrangian follows
\be									  {41}
F_A  =  0
\ee
and
\be									  {42}
G  =  -u(-R)~.
\ee
We get the field equation
\be									  {43}
0  =  -u(-R)^{u-1}R^{ij} - \half g^{ij}(-R)^u +
u\li[(-R)^{u-1}\re]^{;ij}
      - g^{ij}u\square \li[(-R)^{u-1}\re] ~.
\ee
It is enough {}to examine the $00$-component of the field
equation, because all
the other components are fulfilled, if the $00$-component is
fulfilled. We
make the ansatz
\be									  {44}
\dot\ga(t)  =  1 + \gb(t)
\ee
and get
\beq									  {45}
R^{00} & = & -n\dot\gb - 2n\gb - n \nonumber\\
R      & = & -2n\dot\gb -2(n^2 + n)\gb - (n^2 + n) \\
(-R)^m & = & 2nm(n^2 + n)^{m-1}\dot\gb + 2m(n^2 + n)^m\gb +
(n^2 + n)^m~.\nonumber
\eeq
It follows the field equation
\beq									  {46}
0 & = & -2n^2u(u + 1)(n^2 + n)^{u-2}\ddot\gb
        - 2n^3u(u-1)(n^2 + n){u-2}\dot\gb + \nonumber\\
  &   & + nu(2u - n - 1)(n^2 + n)^{u-1}\gb + \li(nu -
\half(n^2 +n)\re)
        (n^2 + n)^{u-1}.
\eeq
Using the condition $2u=n+1$ we get
\be									  {47}
0  =  \ddot\gb + n\dot\gb~.
\ee
All solutions of the linearized field equation are
\be									  {48}
\gb(t)  =  c_1 + c_2e^{-nt}~.
\ee
It follows
\be									  {49}
\ga(t)  =  t + \tilde c_1t + \tilde c_2e^{-nt} + \tilde c_3
\ee
and
\be									 {410}
\lim_{t\to\infty}\frac{\ga(t)}{t}  =  1+\tilde c_1~.
\ee
The $D$-dimensional de Sitter space-time is for the Lagrangian
$F=(-R)^{\frac{D}{2}}$ an attractor solution.

\section{The Lagrangian F=R${\bf\square^k}$R} %5.
\ez

The Lagrangian $(-R)^{\frac{D}{2}}$ leads only {}to a field
equation of fourth
order for $D>2$. The Lagrangian $R\square^k R$ with
$k>0$ give a field equation of higher then fourth order.

For the $00$-component of the field equation we need
\be									  {51}
F_A  =  \square^{k-A}R
\ee
and
\be									  {52}
G  =  2\square^kR
\ee
and get
\beq									  {53}
0 & = & \square^kR\li(2R^{00}-\half R\re) +
2n\dot\ga\square^kR_{,0} +\nonumber\\
  &   & \hspace{2cm} +
\sum_{A=1}^k(\square^{k-A}R)(\square^AR)
                     -
\half(\square^{k-A}R)_{,0}(\square^{A-1}R)_{,0}~.
\eeq
The ansatz
\be									  {54}
\dot\ga(t)  =  1 + \gb(t)
\ee
leads {}to
\be									  {55}
\ba{lcl}
R_{00} & = & -n\dot\gb - 2n\gb -n\\
R      & = & -2n\dot\gb - 2(n^2 + n)\gb - (n^2 + n)~,
\ea
\ee
and
\be									  {56}
\square(\square^kR)  =  (\square^kR)_{,00} +
n(\square^kR)_{,0}~.
\ee
We get the linearized field equation
\be									  {57}
\square^kR  =  (\square^kR)_{,0}~.
\ee
For $k=1$ we have
\be									  {58}
0  =  \gb^{(4)} + 2n\dddot{\gb} + (n^2 - n - 1)\ddot\gb +
(-n^2 - n)\dot\gb
\ee
with the characteristic polynomial
\be									  {59}
P(t)  =  x^4 + 2nx^3 + (n^2 - n - 1)x^2 + (-n^2 - n)x
\ee
with the roots $x_1=1$, $x_2=0$, $x_3=-n$ and $x_4=-n-1$. We
get the
solutions
\be									 {510}
\gb(t)  =  c_1 + c_2e^t + c_3e^{-nt} + c_4e^{-(n+1)t}
\ee
and
\be									 {511}
\ga(t)  =  \tilde c_1t + \tilde c_2e^t + \tilde c_3e^{-nt}
           + \tilde c_4e^{-(n+1)t}
\ee
and
\be									 {512}
\lim_{t\to\infty}\frac{\ga(t)}{t}  =  \infty~.
\ee
The $D$-dimensional de Sitter space-time is for the Lagrangian
$R\square^kR$
not an attractor solution of the field equation. The formula
\be									 {513}
0  =  (\square^{k+1}R)_{,0} - \square^{k+1}R
   =  (\square^kR_{,0} - \square^kR)_{,00} +
n((\square^kR)_{,0} - \square^kR)_{,0}
\ee
for the linearized field equation for $k+1$ leads {}to the
recursive formula for
the characteristic polynomial:
\[
\mbox{characteristic polynomial for~} k+1 =
x(x+n)\cdot\mbox{characteristic polynomial for~} k~.
\]
The characteristic polynomial for k has the roots:
\be									 {514}
\ba{lcll}
x_1 & = & ~~1  & \mbox{simple}\\
x_2 & = & ~~0  & \mbox{k-fold}\\
x_3 & = & -n   & \mbox{k-fold}\\
x_4 & = & -n-1 & \mbox{simple}~.\\
\ea
\ee
We get the solutions
\be									 {515}
\gb(t)  =  S(t) + T(t)e^{-nt} + c_1e^t + c_2e^{(-n-1)t}
\ee
and
\be									 {516}
\ga(t)  =  \widetilde S(t) + \widetilde T(t)e^{-nt} + \tilde
c_1e^t
           + \tilde c_2e^{(-n-1)t}
\ee
with $S,T,\widetilde T$ polynomials at most $k$-th degree and
$\widetilde S$
polynomial at most $k+1$-th degree. For the most solutions is
\be									 {517}
\lim_{t\to\infty}\frac{\ga(t)}{t}  =  \infty
\ee
fulfilled and the de Sitter space-time is not an attractor
solution for the
the field equation derived from Lagrangian $R\square^k R$.

\section{The generalized Lagrangian} %6.
\ez

The results of the last section have shown, that for the
Lagrangian
$R\square^k R$ with $k>1$ the de Sitter space-time is not an
attractor
solution. The Lagrangian $(-R)^{\frac{D}{2}}$ gives only a
fourth order
differential equation. We will try {}to answer the following
question:\\
Are there generalized Lagrangians so, that the de Sitter
space-time is an
attractor solution of the field equation?

First we make the ansatz
\be									  {61}
F  =  \sum_{k=1}^mc_kR\square^kR \qmbox{with} c_m\neq0~.
\ee
In this case is the de Sitter space-time not an attractor
solution, because
for each term is $+1$ a root of the characteristic polynomial
of the linearized
field equation.

Now we make the ansatz
\be									  {62}
F  =  c_0(-R)^{\frac{D}{2}} + \sum_{k=1}^mc_kR\square^kR
\qmbox{with} c_m\neq0~.
\ee
for the generalized Lagrangian. It follows the characteristic
polynomial
\be									  {63}
P(x)  =  x(x+n)\li[c_0 + \sum_{k=1}^mc_kx^{k-1}(x + n)^{k-1}(x
- 1)(x + n - 1)\re]
\ee
for the linearized field equation. The solutions $x_1=0$ and
$x_2=-n$ do not
depend on the coefficients $c_i$ of the Lagrangian. It is
sufficient {}to look
for the roots of the polynomial
\be									  {64}
P(x)  =  c_0 + \sum_{k=1}^mc_kx^{k-1}(x + n)^{k-1}(x - 1)(x +
n - 1)~.
\ee
If the above polynomial only has solutions with negative real
part, then is the
de Sitter space-time an attractor solution for the field
equation. The
transformation
\be									  {65}
z  =  x^2 + nx + \frac{n^2}{4}
\ee
gives
\beq									  {66}
P(x) & = & Q(z) =\nonumber\\
& = & c_0 + \sum_{k=1}^mc_k\li(z - \frac{n^2}{4}\re)^{k-1}
      \li(z - \frac{n^2}{4} - n -1\re)\nonumber\\
& = & c_0 + \sum_{k=1}^mc_k\li(\frac{n^2}{4}\re)^{k-1}
      \li(\frac{n^2}{4} - n - 1\re) +
\sum_{l=1}^{m-1}\Bigg[c_l + \sum_{k=l+1}^m
      c_k\li(-\frac{n^2}{4}\re)^{k-l-1}\cdot\nonumber\\
&   & \cdot\li[{k-1 \choose l-1}\li(-\frac{n^2}{4}\re) - {k-1
\choose l}
      \li(\frac{n^2}{4} + n + 1\re)\re]\Bigg]z^l +
c_mz^m\nonumber\\
& = & d_0 + d_1z + \ldots + d_mz^m~.
\eeq
Now let be
\be									  {67}
\ba{lcll}
a_{ll}\!\! & = & \!\!1  			& l = 0,\ldots,m\\
a_{0k}\!\! & = & \!\!\displaystyle
-\li(-\frac{n^2}{4}\re)^{k-1}
		 \li(\frac{n^2}{4}+n+1\re)	& k = 1,\ldots,m\\
a_{lk}\!\! & = &
\!\!\displaystyle\li(-\frac{n^2}{4}\re)^{k-l-1}\!\!
		 \li[{k-1 \choose l-1}\!\!\li(-\frac{n^2}{4}\re)\!\!
		 - \!\!{k-1 \choose l}\!\!\li(\frac{n^2}{4} + n +
1\re)\re]
						& l < k \le m\\
a_{kl}\!\! & = & \!\!0  			& \mbox{else}~.
\ea
\ee
This gives the equation
\be									  {68}
\li(\ba{c}
d_0 \\ \vdots \\ d_m
\ea\re)  =  A
\li(\ba{c}
c_0 \\ \vdots \\ c_m
\ea\re)\qmbox{with} A \quad\mbox{regular}~.
\ee
The roots of $P(x)$ have a negative real part iff the roots of
$Q(z)$ are from
the set
\be									  {69}
M :=  \li\{x+iy: x > \frac{n^2}{4} \wedge
|y|<n\sqrt{x-\frac{n^2}{4}}\re\}~.
\ee
If the roots $z_k$ of the polynomial $Q(z)$ are elements of
$M$, then the
coefficients $d_k$ are determined by
\be									 {610}
Q(z)  =  \sum_{k=0}^md_kz^k  = \prod_{k=1}^m(z-z_k).
\ee
The coefficients
\be									 {611}
\li(\ba{c}
c_0 \\ \vdots \\ c_m
\ea\re)  =  A^{-1}
\li(\ba{c}
d_0 \\ \vdots \\ d_m
\ea\re)
\ee
belongs {}to a Lagrangian, that gives a field equation with a
de Sitter
attractor solution.
The above considerations have shown that for every $m$ there
exists an
example for coefficients $c_k$, so that the de Sitter
space-time is an
attractor solution for the field equation derived from the
Lagrangian
$c_0(-R)^{\frac{D}{2}} + \suml_{k=1}^mc_kR\square^kR
\qmbox{with} c_m\neq0$.

\section*{Acknowledgements}
\ez

The author would like {}to thank H.-J.~Schmidt and K.~Peters
for discussions
and valuable comments.

\end{document}